\documentclass[twocolumn,aps,showpacs,prl,epsf,graphics,psfig]{revtex4}
\usepackage{graphicx}
\begin{document}
\title{Spin textures in strongly coupled  electron spin  and magnetic or nuclear spin systems in quantum dots}
\author{Ramin M. Abolfath$^{1,2,3,\dagger}$, Marek Korkusinski$^3$, Thomas Brabec$^{2}$, Pawel Hawrylak$^{2,3}$}
\affiliation{
$^{1}$School of Natural Sciences and Mathematics, University of Texas at Dallas, Richardson, TX 75080 \\
$^{2}$University of Ottawa, Physics Department 150 Louis Pasteur, Ottawa, ON, K1N 6N5, Canada \\
$^3$ Institute for Microstructural Sciences, National Research Council of Canada, Ottawa, K1A 0R6 }
\date{\today}

\begin{abstract}
Controlling electron spins strongly coupled to magnetic and nuclear spins in solid state systems is an important challenege in the field of spintronics and quantum computation. We show here  that electron droplets with no net spin  in semiconductor  quantum dots
strongly coupled with magnetic ion/nuclear spin systems break down at low temperature and form a non-trivial anti-ferromagnetic spatially ordered spin-texture of magneto-polarons. The spatially ordered combined electron-magnetic ion spin-texture, associated with spontaneous symmetry-breaking in the parity of electronic charge and spin densities and magnetization of magnetic ions, emerge from ab-initio density functional approach to the electronic system coupled with mean-field approximation for the magnetic/nuclear spin system.
The predicted phase diagram determines the critical temperature as a function of coupling strength and identifies possible phases of the strongly coupled spin system. This prediction may arrest fluctuations in spin system and open the way to control, manipulate and prepare magnetic and nuclear spin ensembles in semiconductor nanostructures.
\end{abstract}
\maketitle


There is currently significant interest in developing quantum information storage and processing capabilities
\cite{Benson2008:qubitbook} using electron (e) and/or hole (h) spins strongly coupled to  spins of either magnetic ions (MI) and/or nuclear spins (NS) in a number of solid state systems. This includes GaAs based gated two-dimensional\cite{Yusa2005:Nature}
and zero-dimensional systems \cite{Petta2005:Science,Reilly2008:PRL,
Tsyplyatev2011:PRL}, InAs self-assembled quantum dots~\cite{Kioseoglou2008:PRL,Makhonin2011:NatMat,Baudin2011:PRL}, CdTe quantum dots~\cite{Besomber2004:PRL,Fernandez-Rossier2004:PRL,Govorov2005:PRB,Qu2005:PRL,Leger2006:PRL,
Qu2006:PRL,Cheng2008:EPL,AbolfathPRL:2007,AbolfathPRL:2008,LeGall2009:PRL,Nguyen2008:PRB,Trojnar2011:PRL,Oszwaldowski2011:PRL}, nanocrystals~\cite{Ochsenbein2009:NN,Viswanatha2011:PRL}, NV centers in diamond \cite{Fuchs2011:NatPhys}, phosphor impurities in silicon~\cite{Kane1998:Nature} and carbon nanotubes~\cite{Churchill2009:NatPhys}.

In these systems electron spins play either a role of qubits or are used to control MIs or NSs.
For e-spin qubits, much effort is directed to determine the role of decoherence by NSs~\cite{Reilly2008:PRL,Tsyplyatev2011:PRL}.
For electron interacting with MI in diluted magnetic semiconductors~\cite{Furdyna1988:JAP,Gaj2011:book}, the central spin
problem has been understood in terms of magneto-polarons (MP) - a cloud of magnetization surrounding a localized e-spin~\cite{Gaj2011:book}. On the other hand, the interaction of many electrons in a spin singlet state with MIs/NSs, e.g., in a metal~\cite{Ruderman1954:PR} or a quantum dot~\cite{Qu2006:PRL,Cheng2008:EPL} induces Ruderman-Kittel-Kasuya-Yosida (RKKY) interaction among MIs/NSs. The question of whether spin textures could form in a strongly coupled two-dimensional electron-NS subsystem has been addressed recently by Loss {\em et al}.~\cite{Loss2011:PRL}. Both  scenarios of MPs and RKKY interactions can be realized in semiconductor quantum dots  containing MIs/NSs by changing electron/hole concentration with a gate, through modulation doping~\cite{Qu2006:PRL,Cheng2008:EPL} or through deformation of quantum dot confining potential~\cite{AbolfathPRL:2008}.
Spin singlet droplets are already employed in initialization of coded qubits in
lateral quantum dots~\cite{Petta2005:Science} and as component of trions in optical manipulation of spin in
self-assembled quantum dots~\cite{Makhonin2011:NatMat,Baudin2011:PRL}.

Currently, magnetic ordering in closed-shell QDs doped with Mn ions is under question. In particular,  recent work by Oszwaldowski {\em et al.}~\cite{Oszwaldowski2011:PRL} suggests that closed-shell QDs doped with Mn do not allow magnetic ordering. The authors came to such conclusion because the ground-state of a nonmagnetic two electron system is a spin singlet but neglected RKKY interaction previously discussed in Ref.~\cite{Qu2006:PRL}.
Hence a full understanding of strongly interacting electronic singlet state with MI/NS in QDs still appears to be missing.
In this work we show that the  closed shell magnetic QDs strongly coupled with magnetic ions  do allow magnetic ordering.
The perturbative RKKY interaction among Mn ions for the closed shell QDs, first discussed in Ref.~\cite{Qu2006:PRL}, does not change the symmetry of electronic ground state. In the present work, our nonperturbative approach predicts novel broken symmetry electronic ground state with spin-textures, a non-trivial form of magnetic ordering, followed by ferromagnetic state of electrons and magnetic ions at very strong coupling.
Because of the degeneracy of the spin-textures under continuous rotational transformation, the magnetization of Mn and the charge and spin densities of e/h are expected to drift over time.
We use ab-initio density functional approach to the e-system and mean-field
approximation for the MIs/NSs. 
We show that beyond RKKY approximation~\cite{Qu2006:PRL} a spin singlet e-droplet corresponding to closed-shell quantum dots strongly couples with MIs/NSs. The e-droplet breaks down and forms a non-trivial anti-ferromagnetic (AFM) spatially ordered spin-texture of MPs at low temperature. For very strongly coupled and weakly confined quantum dot system this transition is followed by a transition to a ferromagnetic state. This prediction opens the way to control, manipulate and prepare MI/NS ensembles in semiconductor nano-structures.

We focus here on  closed-shell QDs~\cite{raymond_studenikin_prl2004} containing $N=2,6,12,\dots$ electrons in the presence of many spins of either NSs or MIs, e.g., mangan in CdTe~\cite{Furdyna1988:JAP,Gaj2011:book,Qu2006:PRL,Cheng2008:EPL}. We approximate many spins by a continuous magnetization~\cite{AbolfathPRL:2007,AbolfathPRL:2008} and study the strongly coupled e-MI system as a function of temperature $T$, electron number $N$, strength of e-MI exchange coupling $J_{sd}$, MI density $n_m$
and strength of confining potential $\omega_0$.
In particular, we find that for a given confining potential, number of  electrons, MI/NS density and their coupling with electrons, there exists a critical temperature, $T^*$ for a two-dimensional nucleation and growth of inhomogeneous AFM spin-textures with broken symmetry. Below $T^*$ and above a critical e-MI exchange coupling strength, spin singlet droplet and a homogeneous magnetization density  breaks down and molecular states of MPs associated with individual e-spins form.
The MPs correspond to the inhomogeneous magnetic field of MIs inducing an effective spatially varying potential  localizing electrons with spin up in different positions from the electrons with spin down, significantly changing electronic spin distribution.
This is to be compared with closed-shell QDs with large confinement potential and  with only two MIs \cite{Qu2006:PRL,Cheng2008:EPL}.
Using exact diagonalization~\cite{Qu2006:PRL} of the interacting e-e and e-MI Hamiltonian it has been shown that the RKKY coupling, a second-order effective interaction between MIs mediated by two e's with opposite spins, describes well the ground state with total magnetic moment $M_z=0$ (AFM ordering) and $M_z=2M$, ferromagnetic (FM)  ordering, depending on the relative position of MIs in a QD, while
maintaining the spin singlet electronic ground state with spin polarization $P_z=0$.
Here $M$ is the total spin of single MI/NS, e.g., $M=5/2$ for Mn.
A related problem has been recently studied where  an analytical variational form of the two electron wave-function
coupled to a large number of MIs, neither two-body singlet nor triplet,  called pseudo-singlet, was introduced to describe partial quantum correlations of the coupled spin singlet-MI system~\cite{Oszwaldowski2011:PRL}.


We use ab-initio density functional to describe the droplet of e's in a parabolic quantum dot with closed electronic shells for electron numbers $N=2,6,12,..$ and mean-field approximation for the MIs/NSs system~\cite{AbolfathPRL:2007}.
We employ spin unrestricted local density approximation (LSDA) for electrons where the many-body Hamiltonian is replaced by the Kohn-Sham (KS) Hamiltonian, $H_{KS}$.
In LSDA, the self-consistent KS orbitals are calculated for spin up and down independently, without any additional symmetrization of their spatial dependence.
The electrons interact with a magnetic field produced by the MIs
which in turn is determined self-consistently by the electron spin density.
The electrons fill KS-orbitals according to Fermi statistics at finite
temperature and the electronic correlations are taken into account via exchange-correlation (XC) energy
functional.
In this approach, the self-consistent solutions form a large class of variational many-body wave-functions, among them the configuration with the lowest free-energy is found.
We also decompose the planar and perpendicular components of the confining potential of a single QD
as described in Refs.~\cite{AbolfathPRL:2007,AbolfathPRL:2008} and expand the electronic wave-functions in terms of its planar, $\psi_{i\sigma}(\vec{r})$, and subband wave function $\xi(z)$, and project $H_{KS}$ onto the XY-plane
by integrating out $\xi(z)$, assuming that only the lowest energy subband is filled.
Hence KS orbitals, $\psi_{i\sigma}(\vec{r})$, are calculated by diagonalizing
$H_{KS} \psi_{i\sigma}(\vec{r}) = \epsilon_{i\sigma}\psi_{i\sigma}(\vec{r})$,
in real-space. Here $\epsilon_{i\sigma}$ is the KS eigen-energies, and
$H_{KS} = \frac{-\hbar^2}{2m^*} \nabla_r^2 + V_\sigma$ with KS effective potential denoted as
$V_\sigma = V_{QD} + V_{H} + V^\sigma_{XC} - \frac{\sigma}{2} h_{sd}(\vec{r})$.
$\hbar$ is the Planck constant, $m^*$ is the e effective mass, $-e$ is the e-charge
and $\sigma=\pm 1$ denotes spin up ($\uparrow$) and down ($\downarrow$).
$V_{QD}$ is the planar confining potential of QD, $V_H$ and $V^\sigma_{XC}$ are Hartree and spin dependent exchange-correlation potentials and
$h_{sd}(\vec{r}) =  J_{em} \int dz |\xi(z)|^2 B_M(M b(\vec{r}, z)/k_BT)$.
Here $B_M(x)$ is the Brillouin function~\cite{Furdyna1988:JAP}, $k_B$ is the Boltzmann constant, and
$b({\bf r}_i)= J_{sd} [n_\uparrow({\bf r}_i) - n_\downarrow({\bf r}_i)]/2$ is the effective field seen by MIs, and
$n_\sigma({\bf r}_i)=\sum_i|\psi_{i\sigma}(\vec{r})\xi(z)|^2 f(\epsilon_{i\sigma})$. Here $f(\epsilon)=1/\{\exp[(\epsilon-\mu)/k_BT] + 1\}$ is the Fermi-Dirac distribution function,
$\mu$ is the chemical potential and $J_{em} = J_{sd} n_m M$ is the mean-field e-Mn or e-NS exchange coupling.
$J_{sd}$ is the local exchange coupling, FM for e-MI and AFM for e-NS or h-MI.
Sign of $J_{sd}$ affects the relative orientations of e/h spin polarization with respect to MIs/NSs.
Finally $n_m$ denotes MI/NS average density.

Our approximation in neglecting spatial distribution of MIs is substantiated by experiments on colloidal nanocrystals and self-assembled QDs~\cite{Ochsenbein2009:NN,Viswanatha2011:PRL} showing that inhomogeneity in the distribution of Mn ions per QD, corresponding up to $\approx 5\%$ doping in typical magnetic semiconductors, can be neglected~\cite{Furdyna1988:JAP}.
For nuclear spins in, e.g., GaAs, inhomogeneity due to different isotopes is unlikely to play an important role.
Typical Mn density is much lower than the NS density, however, because e-Mn exchange coupling is much higher than e-NS exchange coupling, $J_{em}$ for both can be of the same order of magnitude.
Therefore we only present numerical results for a system of interacting e/h and MIs.
For strongly coupled e-MI system we perform numerical calculation for (Cd,Mn)Te where $a^*_B=5.29$ nm, and $Ry^*=12.8$ meV are the
effective Bohr radius and Rydberg, the $sd$ exchange coupling is $J_{sd}=0.015$ eV nm$^3$, effective mass $m^*=0.106$, and
$\epsilon=10.6$~\cite{Qu2005:PRL}.
We consider QD's with electronic shell spacing $\omega_0$ in the range of $ 1-3 Ry^*$,   width of 1 nm, and variable MI density $n_m$.
For definiteness, we focus on the example of Mn isoelectronic impurity in CdTe, with the $z$-component of $\vec{M}_I$ of impurity spin satisfying $M_z=-M, -M+1, \dots, M$ and $M=5/2$.
The direct Mn-Mn AFM coupling is negligible in the range of MI densities considered in this study.

\begin{figure}
\begin{center}\vspace{1cm}
\includegraphics[width=0.9\linewidth]{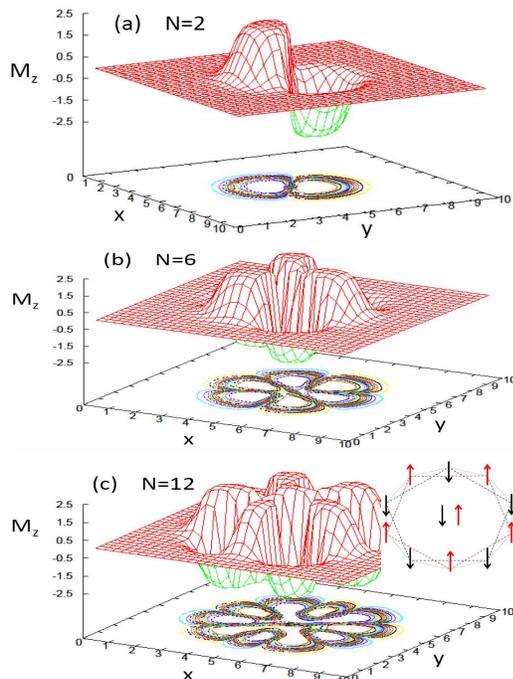} \vspace{-1cm}
\caption{
The spatial profile of  magnetization $M_z$ for a closed shell parabolic QD with $\omega_0=2 Ry^*$ and $n_m=0.1$nm$^{-3}$.
(a), (b), and (c) correspond to $N=2, 6, 12$ at $T=0.5K$ and $J_{sd}=15$ meV nm$^{-3}$.
Coordinates $(x,y)$ are expressed in effective Bohr radius.
Inset in Fig.1c shows a decagon molecular structure of $N=12$ closed-shell QD with a molecular state of 12 MPs, each shown by an arrow.  The direction of arrows show the direction of $M_z$ in each MP.
}
\label{ffig1}
\end{center}
\end{figure}

\begin{figure}
\begin{center}\vspace{1cm}
\includegraphics[width=0.9\linewidth]{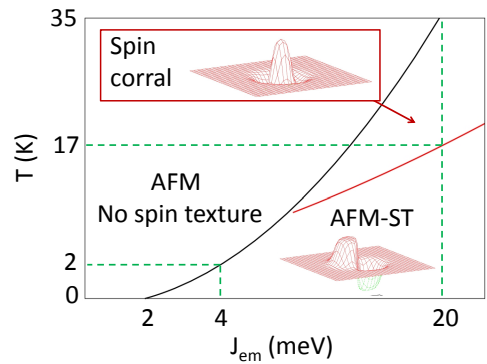} \vspace{-1cm}
\caption{
Phase diagram, temperature $T$ vs e-MI coupling strength $J_{em}$
of $N=2$ closed-shell QD with $\omega_0=2 Ry^*$.
}
\label{ffig3}
\end{center}
\end{figure}

The spatial dependence of the $z$-component of magnetization $M_z(\vec{r})$ of Mn ions, solutions of the self-consistent LSDA equations for the coupled e-MI system, are shown in Fig.~\ref{ffig1}(a-c) for closed shell parabolic QD corresponding to
$\omega_0=2 Ry^*$, $n_m=0.1$ nm$^{-3}$ and temperature $T=0.5$~K for $N=2$, Fig.~\ref{ffig1}(a),  $N=6$, Fig.~\ref{ffig1}(b) and $N=12$, Fig.~\ref{ffig1}(c).
Note that the total magnetization per unit area $A$, $\langle M_z \rangle = \frac{1}{A} \int d^2r \langle M_z(\vec{r})\rangle = 0$, indicating  net AFM ordering of MIs.
These states clearly resemble spin textures with broken rotational symmetry. We see two spatially separated magnetization
clouds for $N=2$ quantum dot, six for $N=6$, and twelve for $N=12$. Hence we consider these states
as molecules of MPs. The inset shows schematically e-spin density for twelve electrons with a   decagon ring structure  and two electrons in the middle in a form of a spin-corral, a circular symmetric spin texture (see Fig.~\ref{ffig3}).
The inhomogeneous spin density gives way to  a uniform solution in which $M_z=s_z=0$ for $T>T^* \approx 2$ K
independent of $N$.
Note that this finding is in contrast with open shell FM states that are stable up to
$T\approx 20$K~\cite{AbolfathPRL:2007,AbolfathPRL:2008} and the FM state proposed recently
in Ref.~\cite{Oszwaldowski2011:PRL} for closed-shell QDs.

\begin{figure}
\begin{center}\vspace{1cm}
\includegraphics[width=0.8\linewidth]{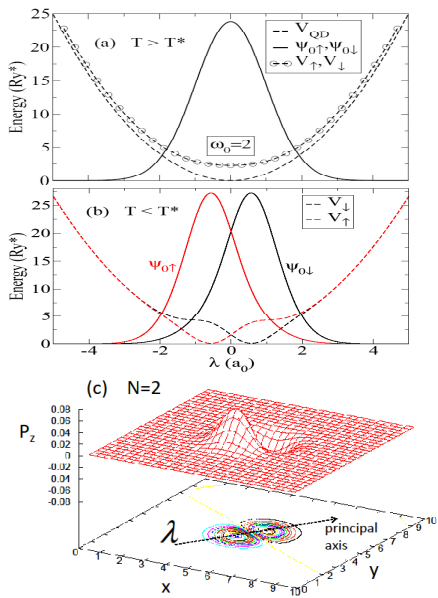}\\ \vspace{-1.0cm}
\caption{
The spatial dependence of wavefunctions $\psi_{0\uparrow} ,  \psi_{0\downarrow}$, effective potentials and $P_z$ in a QD with $N=2$,
$\omega_0=2 Ry^*$, $n_m=0.1$nm$^{-3}$.
In (a) $T > T^*$, $M_z=P_z=0$ and $\psi_{0\uparrow} = \psi_{0\downarrow}$. The dashed lines show the QD
parabolic confining potential, $V_{\rm QD}$. The spin-dependent effective potential
$V_\sigma = V_{\rm QD} + V_H + V^\sigma_{XC} - \sigma h_{sd}/2$ is shown by dashed lines with open circles, with
$V_\uparrow = V_\downarrow$.
In (b) $T < T^*$. A pronounced separation of spin up and down wavefunction and two asymmetric bumps in  $V_\uparrow$ and $V_\downarrow$ are shown. In (b) the bottom of potentials
$V_\uparrow$ and $V_\downarrow$ are shifted to zero.
The horizontal-axis used in (a) and (b) is plotted along the diagonal-direction
of $P_z$ shown as principal  axis by a dashed-arrow in (c), and denoted by $\lambda$ in (b).
Planar coordinates $(x,y)$ are expressed in effective Bohr radius ($a_0$).
}
\label{ffig2}
\end{center}
\end{figure}

To investigate the origin and stability of AFM states presented in this work, we focus on a closed-shell QD
with $N=2$. Let us first consider the RKKY interaction between two MIs in a QD~\cite{Qu2006:PRL} with $N=2$ at positions $\vec{R_1}=(X,0)$ and $\vec{R_2}=(-X,0)$:
$J(X)=-\gamma (2-5X^2)e^{-X^2}$ is the e-mediated effective interaction between two MIs, where $\gamma= (J^{2D}/\pi l_0)^2\times(1/(16 \omega_0))$ and distance is measured in $l_0=\sqrt{\hbar/(2m^* \omega_0)}$. Here $J^{2D}=J_{sd} 2/d$ and $d$ is the thickness of QD in perpendicular direction~\cite{Qu2006:PRL}.
We see that for
$X>\sqrt{2/5}$ the interaction is AFM. It is now possible to imagine that the RKKY interaction triggers a broken symmetry state where the exchange interaction localizes an electron with spin up on the left MI and spin down electron on the right MI, a two-atomic molecule of MP, seen in Fig.~\ref{ffig1}(a).
In Figs.~\ref{ffig2} we show self-consistent solutions of LSDA equations including the effective potential $V_\sigma$ and spatial profile of spin-dependent KS wave-functions for parabolic confining potential with level spacing $\omega_0=2 Ry^*$.
In Fig.~\ref{ffig2}(a), $T=2$K and the paramagnetic state with $M_z=P_z=0$ is the ground state.
The spin-dependent effective potential $V_\sigma$ is shown with open-circles in which
$V_\uparrow = V_\downarrow$. In this case we find that the self-consistent KS eigen-energies and
eigen-states with  opposite spins are identical, $\epsilon_{0\uparrow} = \epsilon_{0\downarrow}$ and
$\psi_{0\uparrow} = \psi_{0\downarrow}$.
In Fig.~\ref{ffig2}(b) we present our results for $T=0.5K$.
Interestingly, with decreasing temperature a continuous displacement in spin-dependent wave-functions develops that
spontaneously breaks the circular symmetry of the QD as discussed above.
In this case we find two degenerate spatially
separated wavefunctions $\psi_{0\uparrow}$ and $\psi_{0\downarrow}$ with $\epsilon_{0\uparrow} = \epsilon_{0\downarrow}$.
This is analogues to the nucleation of para-H$_2$ molecule from para-He if two nuclei of He undergo a
continuous fragmentation.
A two-dimensional profile of e-spin polarization $P_z=(n_\uparrow - n_\downarrow)/2$ shown in (c) is also
a result of spontaneous spatial spin-separation of KS orbitals.
Two asymmetric bumps observed in effective potentials $V_\uparrow$ and $V_\downarrow$ are the result of spontaneous formation of AFM spin texture in $M_z$, shown explicitly in Fig.\ref{ffig1}.

The spontaneous symmetry breaking and the formation of MPs depends on the parameters of the system. For example, in the example studied here, the critical temperature depends on e-MI exchange coupling strength $J_{em}= J_{sd} n_m M$, which is proportional to exchange coupling, MI density and MI spin. Figure~\ref{ffig3} shows the phase diagram of $N=2$ closed shell QD with $\omega_0=2 Ry^*$ and $n_m=0.1$ nm$^{-3}$. The curves show critical temperature as a function of $J_{em}$. In the limit of low $J_{em}$, e.g., either for e and/or low density of MIs, a direct transition from AFM spin texture, shown as inset, to normal state is seen. For high $J_{em}$, e.g., for either h and/or high density of MIs, a transition through an intermediate state of spin-corral, shown as inset  (see Fig.~\ref{ffig3}) is observed. This is a transition from rotational symmetry breaking state, stable at low temperatures, to another type of spin texture with rotational symmetry, stable within $T^*_1 \leq T \leq T^*_2$. For $J_{sd}=75$ meV nm$^3$, corresponding to h-MI exchange coupling, and $n_m=0.1$ nm$^{-3}$ we find $T^*_1=17$ and $T^*_2=35$ K.
We emphasize that the total spin of electrons is zero for the range of parameters considered in this study. However, by increasing $J_{em}$, above a critical value that is determined by the energy difference in the lowest unoccupied and highest occupied KS energy levels, a transition to spin-triplet takes place~\cite{AbolfathPRL:2007,Nguyen2008:PRB,AbolfathPRL:2008}.
For the typical quantum dots investigated here, the critical value of $J^*_{em} \approx 30$ meV is much higher than the range of exchange  couplings $J_{em} $ responsible for the  spin textures predicted here.


To summarize, low temperature solutions of ab-initio density functional approach to  electrons in closed-shell quantum dots strongly coupled with the magnetic and/or nuclear spins described in the mean-field approximation show the existence of a broken-symmetry spatially inhomogeneous state
of geometrically ordered molecules of magneto-polarons.
The phase diagram of topologically stable non-trivial AFM states determined by the number $N$ of electrons in a quantum dot, MI density, exchange coupling strength, temperature and confining potential is predicted. The predicted  ordered spin states may arrest fluctuations in the spin system and open the way to control, manipulate and prepare MI/NS ensembles in semiconductor nanostructures for quantum information processing and storage.

Acknowledgement: The authors thank the NSERC, QuantumWorks and CIFAR for support.
RMA thanks the support from Texas Advanced Computing Center (TACC) for computer resources.


$\dagger$ Present address: Dept. of Therapeutic Radiology, Yale School of Medicine, Yale University, New Haven, CT 06520


\begin{thebibliography}{99}



\bibitem{Benson2008:qubitbook}
{\em Semiconductor quantum bits}, World Scientific, 2008; O. Benson and F. Henneberger, Editors.



\bibitem{Yusa2005:Nature}
G. Yusa {\em et al.},
Nature 434, 1001 (2005).







\bibitem{Petta2005:Science}
J. R. Petta {\em et al.},
Science {\bf 309}, 2180 (2005).


\bibitem{Reilly2008:PRL}
D. J. Reilly {\em et al.},
Phys. Rev. Lett. {\bf 104}, 236802 (2010).

\bibitem{Tsyplyatev2011:PRL}
O. Tsyplyatev and D. Loss, Phys. Rev. Lett. {\bf 106}, 106803 (2011).
A. V. Khaetskii {\em et al.},
Phys. Rev. Lett. {\bf 88}, 186802 (2002);
R. de Sousa and S. Das Sarma, Phys. Rev. B {\bf 67}, 033301 (2003);
W. Yao {\em et al.},
Phys. Rev. B {\bf 74}, 195301 (2006);
G. Chen {\em et al.},
Phys. Rev. B {\bf 76}, 045312 (2007).



\bibitem{Kioseoglou2008:PRL}
G. Kioseoglou {\em et al.},
Phys. Rev. Lett. {\bf 101}, 227203 (2008).

\bibitem{Makhonin2011:NatMat}
M. N. Makhonin {\em et al.},
Nature Materials {\bf 10}, 844 (2011).

\bibitem{Baudin2011:PRL}
E. Baudin {\em et al.},
\prl {\bf 107}, 197402 (2011).




\bibitem{Besomber2004:PRL}
L. Besombes {\em et al.},
Phys. Rev. Lett. {\bf 93}, 207403 (2004).

\bibitem{Fernandez-Rossier2004:PRL}
J. Fern\'andez-Rossier and L. Brey, Phys. Rev. Lett. {\bf 93}, 117201 (2004).

\bibitem{Govorov2005:PRB}
A. O. Govorov, Phys. Rev. B {\bf 72}, 075358 (2005).

\bibitem{Qu2005:PRL}
F. Qu and P. Hawrylak, Phys. Rev. Lett. {\bf 95}, 217206 (2005).

\bibitem{Leger2006:PRL}
Y. L\'eger {\em et al.},
Phys. Rev. Lett. {\bf 95}, 047403 (2005);
Y. L\'eger {\em et al.},
Phys. Rev. Lett. {\bf 97}, 107401 (2006).



\bibitem{Qu2006:PRL}
F. Qu and P. Hawrylak, Phys. Rev. Lett. {\bf 96}, 157201 (2006).

\bibitem{Cheng2008:EPL}
S.-J. Cheng  and P. Hawrylak,
Eur. Phys. Lett. {\bf 81}, 37005 (2008).



\bibitem{AbolfathPRL:2007}
R. M. Abolfath {\em et al.},
Phys. Rev. Lett. {\bf 98}, 207203 (2007);
New J. Phys. {\bf 9}, 353 (2007).



\bibitem{AbolfathPRL:2008}
R. M. Abolfath {\em et al.},
Phys. Rev. Lett. {\bf 101}, 207202 (2008).



\bibitem{LeGall2009:PRL}
C. Le Gall {\em et al.},
\prl {\bf 102}, 127402 (2009).

\bibitem{Nguyen2008:PRB}
Nga T. T. Nguyen and F. M. Peeters,
Phys. Rev. B {\bf 78}, 045321 (2008);
Phys. Rev. B {\bf 80}, 115335 (2009);

\bibitem{Trojnar2011:PRL}
A. Trojnar {\em et al.},
Phys. Rev. Lett. {\bf 107}, 207403 (2011).


\bibitem{Oszwaldowski2011:PRL}
R. Oszwaldowski {\em et al.},
Phys. Rev. Lett. {\bf 106}, 177201 (2011).


\bibitem{Ochsenbein2009:NN}
S. T. Ochsenbein {\em et al.},
Nature Nanotechnology {\bf 4}, 681 (2009);
R. Beaulac {\em et al.},
Phys. Rev. B {\bf 84}, 195324 (2011).
K. M. Whitaker {\em et al.},
Nano Letters {\bf 11}, 3355 (2011).

\bibitem{Viswanatha2011:PRL}
R. Viswanatha {\em et al.},
Phys. Rev. Lett. {\bf 107}, 067402 (2011).





\bibitem{Fuchs2011:NatPhys}
G. D. Fuchs {\em et al.},
Nature Physics {\bf 7}, 789 (2011);



\bibitem{Kane1998:Nature}
B. E. Kane, Nature {\bf 393}, 133 (1998).



\bibitem{Churchill2009:NatPhys}
H. O. H. Churchill {\em et al.},
Nature Physics {\bf 5}, 321 (2009).


\bibitem{Furdyna1988:JAP}
J. K. Furdyna, J. Appl. Phys. {\bf 64}, R29 (1988);
T. Dietl {\em et al.},
Phys. Rev. B {\bf 63}, 195205 (2001);
M. Abolfath {\em et al.}, \prb {\bf 63}, 054418 (2001).

\bibitem{Gaj2011:book}
{\em The Physics of Diluted Magnetic Semiconductors}, Springer series in materials science,
 Springer-Verlag 2011, J. Gaj and J. Kossut, Editors.



\bibitem{Ruderman1954:PR}
M. A. Ruderman and C. Kittel, Phys. Rev. {\bf 96}, 99 (1954); T. Kasuya, Prog. Theor. Phys. {\bf 16}, 45 (1956); K. Yosida, Phys. Rev. 106, 893 (1957).

\bibitem{Loss2011:PRL}
D. Loss {\em et al.},
Phys. Rev. Lett. {\bf 107}, 107201 (2011).




\bibitem{raymond_studenikin_prl2004}
S. Raymond {\em et al.},
Phys. Rev. Lett.  {\bf 92}, 187402 (2004).



\end{thebibliography}
\end{document}